\documentclass[12pt]{iopart}

\expandafter\let\csname equation*\endcsname\relax
\expandafter\let\csname endequation*\endcsname\relax

\usepackage{graphicx}
\usepackage{mathrsfs}
\usepackage{graphics}
\usepackage{amsmath}
\usepackage{color}
\usepackage[version=3]{mhchem}
\usepackage[format=plain,labelsep=period]{caption}
\makeatletter

\renewcommand{\maketag@@@}[1]{\hbox{\m@th\normalsize\normalfont#1}}%

\makeatother

\begin{document}

\title{Thermal dissipation of the quantum spin Hall edge states in HgTe/CdTe quantum well}

\author{Jing-Yun Fang$^{1,2}$, Yu-Chen Zhuang$^{1,2}$, Ai-Min Guo$^{3}$ and Qing-Feng Sun$^{1,2,4,*}$}

\address{$^1$ International Center for Quantum Materials, School of Physics, Peking University, Beijing 100871, China}
\address{$^2$ Hefei National Laboratory, Hefei 230088, China}
\address{$^3$ Hunan Key Laboratory for Super-microstructure and Ultrafast Process, School of Physics and Electronics, Central South University, Changsha 410083, China}
\address{$^4$ Beijing Academy of Quantum Information Sciences, West Bld.\#3,No.10 Xibeiwang East Rd., Haidian District, Beijing 100193, China}
\ead{sunqf@pku.edu.cn}
\vspace{10pt}
\begin{indented}
\item[]{\today}
\end{indented}

\begin{abstract}
Quantum spin Hall effect is characterized by topologically protected helical edge states. Here we study the thermal dissipation of helical edge states by considering two types of dissipation sources. The results show that the helical edge states are dissipationless for normal dissipation sources with or without Rashba spin-orbit coupling in the system, but they are dissipative for spin dissipation sources. Further studies on the energy distribution show that electrons with spin-up and spin-down are both in their own equilibrium without dissipation sources. Spin dissipation sources can couple the two subsystems together to induce voltage drop and non-equilibrium distribution, leading to thermal dissipation, while normal dissipation sources cannot. With the increase of thermal dissipation, the subsystems of electrons with spin-up and spin-down evolve from non-equilibrium finally to mutual equilibrium. In addition, the effects of disorder on thermal dissipation are also discussed. Our work provides clues to reduce thermal dissipation in the quantum spin Hall systems.
\end{abstract}


\section{Introduction}
In recent years, topological phenomena have attracted much attention both in condensed matter physics and material science, generating significant influence in fundamental and applied research~\cite{Rev.Mod.Phys.83.1057,Rev.Mod.Phys.92.021003}. Topological insulator, a new topologically nontrivial quantum material, has made great theoretical and experimental research progress~\cite{Rev.Mod.Phys.82.3045,JPCM.34.405302}. Owing to its unique band structure with gapless edge (or surface) states surviving in the band gap of insulating bulk, the topological insulator  can behave as an insulator inside and a metal on the edge (or surface). This special band structure leads to many interesting physical properties, one of the most famous is the quantum spin Hall effect (QSHE)~\cite{PRL.96.106802,Science.346.1344,PRB.103.085109}, which was first proposed by Kane and Mele based on graphene~\cite{PRL.95.226801,PRL.95.146802} and later confirmed by Zhang in
an inverted HgTe/CdTe quantum well~\cite{Science.314.1757,Science.318.766}. Due to the time-reversal symmetry (TRS), a pair of topologically protected
edge states (known as helical edge states) counterpropagates along a given edge of the quantum spin Hall insulator (QSHI). Such two edge states, characterized by $Z_2$ topological invariant~\cite{PRL.95.146802}, are spin-momentum locked
with opposite spin polarization~\cite{NaturePhys.8.486}, behaving as a typical characteristic of the QSHE. Because of the helical edge states, the QSHE can produce a quantized conductance of $2e^2/h$. In principle, the helical edge states are robust against backscattering, thus can be served as dissipationless conducting channels~\cite{Science.325.294,Science.301.1348}. On this basis, one can design electronic devices with less energy dissipation by the QSHIs, making QSHIs attractive for material science. Currently, various QSHIs have been discovered successively~\cite{PRL.107.136603,PRL.114.096802,JPCM.34.245301,PRL.107.076802,PRL.111.136804,Nat.Phys.13.683}, such as InAs/GaSb quantum well~\cite{PRL.107.136603,PRL.114.096802,JPCM.34.245301}, Silicene~\cite{PRL.107.076802}, Tin films~\cite{PRL.111.136804}, monolayer \ce{WTe2}~\cite{Nat.Phys.13.683}, etc.

With the discovery of a large number of QSHIs,
research on the helical edge states has attracted increasing attention, such as the penetration depth of edge states~\cite{PRB.83.121310}, p-n junction~\cite{JPCM.26.085301}, finite size effect~\cite{PRL.101.246807}, robustness against the magnetic field~\cite{anref1}, and so on. Although theoretical studies have shown that
QSHE is robust because of topologically protected helical edge states, it is fundamentally different from the integer quantum Hall effect (IQHE)
in terms of the robustness of topological protection.
In IQHE, the edge states are chiral, which propagate along the edge of the sample unidirectionally and can be realized macroscopically. The quantized Hall conductance in IQHE is very accurate and independent of the specific details of materials, thus it can even be used to calibrate the resistance unit and determine the fine-structure constant~\cite{PRL.45.494,PRB.23.5632,PRL.49.405}. However, the quantized conductance in QSHE is usually found to
deviate from expected value of $2e^2/h$ in real experimental measurements~\cite{Science.359.76,PRL.112.026602}. In particular, the quantized conductance plateau will be destroyed severely when the sample becomes larger and reaches a few microns~\cite{PRL.103.036803}. Several possible mechanisms have been proposed to explain this phenomenon of quantized plateau breaking, such as nonmagnetic impurities~\cite{PRL.122.016601}, coupling to phonons~\cite{PRL.108.086602}, charge puddles~\cite{PRL.110.216402}, etc.,
which are essential because the topologically protected helical edge states against backscattering are compromised.

When backscattering occurs at the edge states,
a non-zero resistance is usually generated~\cite{PRB.90.115309,PRB.103.235164}, and this is associated with the occurrence of dissipation and the production of entropy. Dissipation caused by electron transport is often irreversible, because ordered electrical energy will be transferred to disordered thermal energy through electron-phonon scattering. This will raise the systemic temperature. Too much heat production and too high temperature are one of
the main problems that hinder the development of electronic devices. Therefore, studying the microscopic mechanism of energy dissipation
is of fundamental importance for the application of electronic devices. Previous research of thermal dissipation in IQHE has made some progress~\cite{PRR.2.013337,Science.358.1303}, and a counterintuitive phenomenon that thermal dissipation occurs
along the chiral edge states has been found~\cite{Nature.575.628,PRB.104.115411,newref1}.
As we mentioned before, IQHE is more robust than QSHE~\cite{PRL.49.405,PRL.103.036803}.
Compared to the chiral edge states in IQHE,
the helical edge states in QSHE should be more likely to cause inelastic scattering and thus induce dissipation, because the counterpropagating helical edge states are easier to destroy than the unidirectionally propagating chiral edge states. However, so far there has been little research on thermal dissipation in QSHE~\cite{PRL.108.086602,PRL.105.146103}. In view of this situation, it is desirable to explore the thermal dissipation in QSHE.

In this paper, based on an inverted HgTe/CdTe quantum well, we investigate the thermal dissipation of helical edge states in the presence of dissipation sources. In real samples, dissipation sources inevitably exist. Considering the spin index, we can divide them into two main categories. One is the normal dissipation sources which make electrons lose energy to the environment in the form of heat energy but maintain the spin memory, and the other is the spin dissipation sources which make electrons lose both energy and spin memory. The former are caused by electron-phonon interaction~\cite{PRL.108.086602}, and the latter are ascribed to the combination of lattice vibration (phonon) and magnetic impurities~\cite{PRL.106.236402,PRL.110.206803},
embedded nuclear spin~\cite{PRB.96.081405,PRB.97.125432}, spin-orbit coupling (SOC)~\cite{PRL.129.197201,NatureMater.9.259,JPCM.25.335503}, etc. In our calculations, dissipation sources are simulated by B\"{u}ttiker's virtual leads~\cite{PRL.57.1761}. Through the transmission of electrons between the central region
in the real system and the virtual leads,
energy exchange and thermal dissipation can be realized. By applying the Landauer-B\"{u}ttiker formalism combined with the non-equilibrium Green's function, the local heat generation caused by thermal dissipation and the energy distribution are calculated. The results show that no thermal dissipation occurs for normal dissipation sources with or without Rashba SOC.
However, thermal dissipation occurs along the helical edge states for spin dissipation sources.
Electrons in the helical edge states with spin-up and spin-down are individually in equilibrium at the beginning, but they are not equilibrium with each other. When thermal dissipation occurs, electrons at the higher energy will relax to the lower energy, and the distribution will finally
tend to a new equilibrium with increasing dissipation strength. Besides, we also analyze the effects of disorder on thermal dissipation.
For normal dissipation sources, disorder has little effect on thermal dissipation as long as the helical edge states are preserved,
but it will enhance the thermal dissipation for spin dissipation sources.

The rest of the paper is organized as follows.
In section~2, we describe the model of the system and give the details of our calculations. In section~3,
we show the numerical results and some discussions.
Finally, a brief conclusion is presented in section~4.

\begin{figure}
	\includegraphics[width=0.8\textwidth]{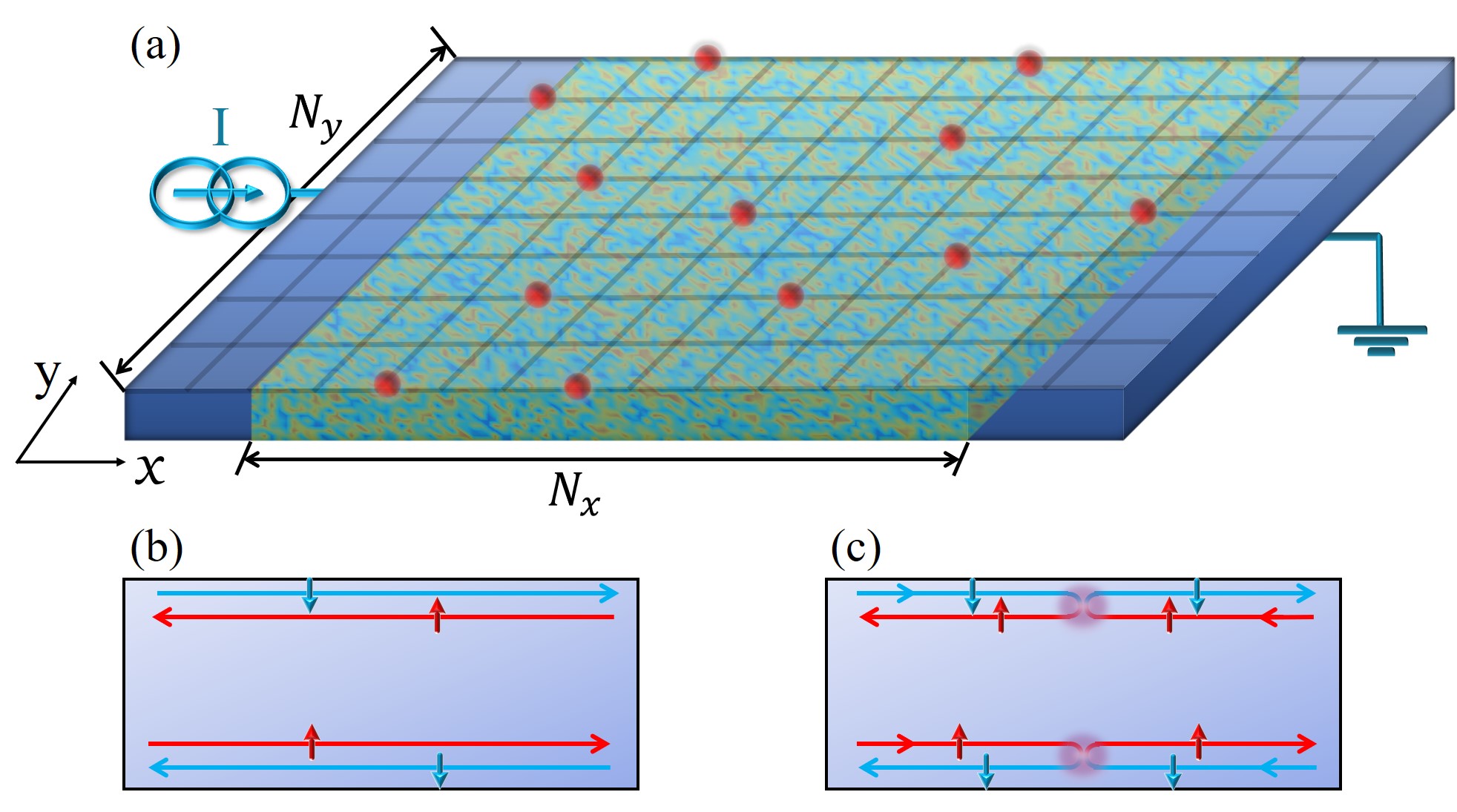}
	\centering
	\caption{(a) Schematic diagram for a two-terminal device based on disordered HgTe/CdTe quantum well. Here the red balls represent dissipation sources. In this diagram, the sizes of device are $N_y=9$ and $N_x=12$. (b) Spin-polarized helical edge states in QSHE, spin-up and spin-down channels propagate in opposite directions. (c) Mixing of spin-up and spin-down channels due to backscattering.}
	\label {fig:1}
\end{figure}

\section{Model and method}
First of all, we introduce the effective Hamiltonian of the HgTe/CdTe quantum well with Anderson impurity, Rashba SOC, and dissipation sources.
Its discrete form in the tight-binding representation of a square lattice
(see figure~\ref{fig:1}(a)) can be written as~\cite{PRB.85.125401,anref2}
\begin{eqnarray} \label{eq:1}
H=\sum_{\bf i}[\varphi_{\bf i}^{\dag}T_0\varphi_{\bf i}+(\varphi_{\bf i}^{\dag}T_{\hat{\bf x}}\varphi_{\bf i+\hat{x}}+\varphi_{\bf i}^{\dag}T_{\hat{\bf y}}\varphi_{\bf i+\hat{y}}+\textrm{H.c.})]+H_d,
\end{eqnarray}
where $H_d$ is
\begin{eqnarray}\label{eq:2}
H_d = \sum_{{\bf i},k}[\epsilon_{k}\psi_{{\bf i}k}^{\dag}\psi_{{\bf i}k}+(t_{k}\psi_{{\bf i}k}^{\dag}\varphi_{\bf i}+\textrm{H.c.})].
\end{eqnarray}
Here ${\bf i}=(x,y)$ is the site index, and $\hat{\bf x}$ and $\hat{\bf y}$ are, respectively, the unit vectors along $x$ and $y$ directions. $\varphi_{\bf i}=(a_{\bf i},b_{\bf i},c_{\bf i},d_{\bf i})^T$ represents the four annihilation operators of electron on the site {\bf i} at the special states $|s,\uparrow\rangle$, $|p_x+ip_y,\uparrow\rangle$, $|s,\downarrow\rangle$, and $|-(p_x-ip_y),\downarrow\rangle$, respectively. $T_0=\textrm{diag}(E_s+W_{\bf i},E_p+W_{\bf i},E_s+W_{\bf i},E_p+W_{\bf i})$ is the matrix of on-site energy with $E_{s/p}=C\pm M-4(D\pm B)/a^2$. $W_{\bf i}$ is the disordered on-site energy, which is uniformly distributed in the range $[-\frac{W}{2},\frac{W}{2}]$ with the disorder strength $W$. The hopping matrices along $x$ and $y$ directions are, respectively,

\begin{eqnarray}\label{eq:3}	
  T_{\hat{\bf x}}&=&
  \left(
    \begin{array}{cccc}
      V_{ss} & V_{sp} & V_{R} & 0 \\
      -V_{sp}^{*} & V_{pp} & 0 & 0 \\
      -V_{R}^{*} & 0 & V_{ss} & V_{sp}^{*} \\
      0 & 0 & -V_{sp} & V_{pp} \\
    \end{array}
  \right), \nonumber  \\
  T_{\hat{\bf y}}&=& \left(
      \begin{array}{cccc}
        V_{ss} & iV_{sp} & -iV_{R} & 0 \\
        iV_{sp}^{*} & V_{pp} & 0 & 0 \\
        -iV_{R}^{*} & 0 & V_{ss} & -iV_{sp}^{*} \\
        0 & 0 & -iV_{sp} & V_{pp} \\
    \end{array}
  \right),
\end{eqnarray}
with $V_{ss/pp}=(D\pm B)/a^2$, $V_{sp}=-iA/(2a)$ and $V_R=\alpha/(2a)$. Here, $a$ is the lattice constant, $\alpha$ is the strength of Rashba SOC. $A$, $B$, $C$, $D$, and $M$ are the system parameters, which can be experimentally controlled~\cite{Science.318.766}. $H_d$ in equation~(\ref{eq:2}) represents
the Hamiltonian of virtual leads and their couplings to central sites, which is used to simulate the dissipation sources.
$\psi_{{\bf i}k}^\dag$ ($\psi_{{\bf i}k}$) is the creation (annihilation) operator for electrons in the virtual lead ${\bf i}$, $t_{k}$ is the coupling strength,
and momentum $k$ labels continuous states in each virtual lead ${\bf i}$.
Here the virtual lead ${\bf i}$ just couples with the site ${\bf i}$ in the central region, so both the sites and the virtual leads can be described
by the same notation ${\bf i}$. Besides, we set that the sites in the central region are randomly selected to be coupled with the virtual lead by a probability of $\eta$.
The left and right leads are also composed of HgTe/CdTe quantum well, but without on-site disorder and dissipation sources (see figure~\ref{fig:1}(a)).

By using the Landauer-B\"{u}ttiker formula, the electric current and the current-induced local heat generation in lead $p$ (either real or virtual lead) with spin $\sigma$ can be expressed as \cite{Datta}
\begin{eqnarray}\label{eq:4}
  I_{p\sigma}&=& \frac{e^2}{h}\sum_{q\ne p,\sigma'}T_{p\sigma,q\sigma'}
  \left(V_{p\sigma}-V_{q\sigma'}\right), \nonumber  \\
  Q_{p\sigma}&=&-\frac{e^2}{h}\sum_{q\ne p,\sigma'}T_{p\sigma,q\sigma'}
  \left(V_{p\sigma}V_{q\sigma'}-\frac{1}{2}V_{p\sigma}^2-\frac{1}{2}V_{q\sigma'}^2\right).
\end{eqnarray}
Here $V_{p\sigma}$ is the spin-dependent bias in lead $p$. The transmission coefficient from lead $q$ with spin $\sigma'$ to lead $p$ with spin $\sigma$ is
\begin{equation}\label{eq:5}
T_{p\sigma,q\sigma'}=\textmd{Tr}[{\bf \Gamma}_{p\sigma}{\bf G}^r{\bf \Gamma}_{q\sigma'}{\bf G}^a],
\end{equation}
where the linewidth function ${\bf \Gamma}_{p\sigma}$ and the Green's function ${\bf G}^r$ are
\begin{eqnarray}\label{eq:6}
{\bf \Gamma}_{p\sigma}&=&i({\bf \Sigma}_{p\sigma}^r-{\bf \Sigma}_{p\sigma}^{a}), \nonumber  \\
{\bf G}^r&=&[{\bf G}^a]^\dag=[E_F {\bf I}-{\bf  H}_{\rm cen}-\sum_{p\sigma}{\bf \Sigma}_{p\sigma}^r]^{-1}.
\end{eqnarray}\\
In equation~(\ref{eq:6}), ${\bf H}_{\rm cen}$ is the Hamiltonian of the central region, and ${\bf \Sigma}_{p}^r=[{\bf \Sigma}_{p}^{a}]^\dagger$ is the retarded self-energy due to the coupling to lead $p$. For the real lead $p$, the self-energy ${\bf \Sigma}_{p}^r$ can be calculated numerically~\cite{Sancho}. For the virtual lead $p$, ${\bf \Sigma}_{p}^r=-\frac{i}{2}\Gamma$ with $\Gamma=2\pi\rho|t_k|^2$ the dissipation strength and $\rho$ the density of states of the virtual lead.

In our calculations, a small bias $V_{d}$ is applied between left and right leads with $V_{L\uparrow}=V_{L\downarrow}=-V_{R\uparrow}=-V_{R\downarrow}=\frac{V_{d}}{2}$, which drives a current $I=I_L=-I_R=I_{L\uparrow}+I_{L\downarrow}$ flowing along the longitudinal direction. For normal dissipation sources, spin-up and spin-down channels are independent of each other. In this case, electrons will exchange energy with the environment (virtual leads) but maintain the spin memory when they go into and then come back from the virtual leads. Therefore, the number of electrons for spin-up and spin-down is separately conserved. Then, for each virtual lead ${\bf i}$, the current has
the constraint that $I_{{\bf i} \uparrow}=I_{{\bf i}\downarrow}=0$, and $V_{{\bf i}\uparrow}$ is usually not equal to $V_{{\bf i}\downarrow}$. But for spin dissipation sources, the spin-up and spin-down channels will couple together and spin-flip backscattering will occur. Therefore, electrons will also lose their spin memories when they exchange energy with the environment. However, the total number of electrons is still conserved. So one have $V_{{\bf i}\uparrow}=V_{{\bf i}\downarrow}$
and $I_{{\bf i}\uparrow}+I_{{\bf i}\downarrow}=0$ for
each virtual lead ${\bf i}$.
Combining equation~(\ref{eq:4}) with all the above boundary conditions,
the rest voltage $V_{p\sigma}$, the current $I_{p\sigma}$, and the local heat generation $Q_{p\sigma}$
in each lead $p$ with spin $\sigma$ can be obtained (see Appendix A for the detailed calculation process).
Then the two-terminal conductance $G=\frac{e^2}{h}\sum_{q\ne L,\sigma\sigma'}T_{L\sigma,q\sigma'}(V_{L\sigma}-V_{q\sigma'})/V_{d}$ and local heat generation $Q_{\bf i}(x,y)=\sum_{\sigma}Q_{\bf i\sigma}$ at position ${\bf i}=(x,y)$ can be calculated and will be presented next.

In the following numerical calculations,
the five material parameters for a realistic HgTe/CdTe quantum well are set as $A=364.5$ meV$\cdot$nm, $B=-686$ meV$\cdot\textrm{nm}^2$, $C=0$, $D=-512$ meV$\cdot\textrm{nm}^2$, and $M=-10$ meV. These model parameters derived from experimental data guarantee
that the energy band of the HgTe/CdTe quantum well is
inverted~\cite{Science.318.766}. The lattice constant is adopted as $a=3$ nm. The width of the central region is $N_ya=240$ nm with $N_y=80$ and the length of the central region is $N_xa=450$ nm with $N_x=150$. The probability of virtual leads in the central region is $\eta=10\%$. When calculating the local heat generation $Q_{\bf i}$, positions of dissipation sources are averaged over 200 times.

\begin{figure}
  \includegraphics[width=0.8\textwidth]{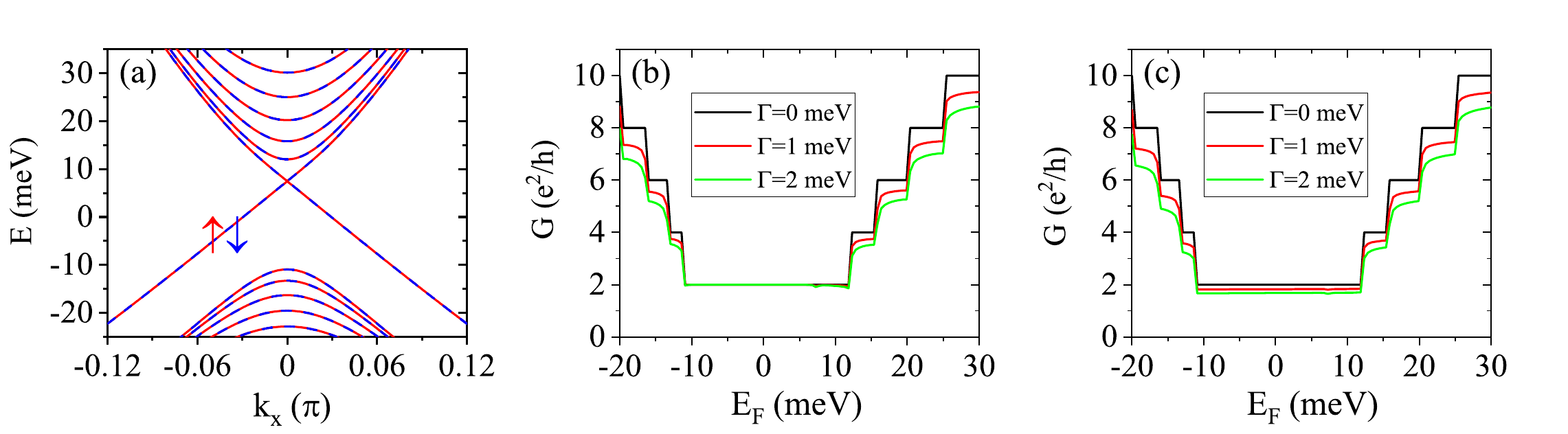}
  \centering
  \caption{(a) Energy band of the HgTe/CdTe quantum well with the width $N=80$ and parameter $M=-10$ meV, in which spin-up (red lines) and spin-down (blue lines) are degenerate. (b) and (c) are conductance $G$ versus Fermi energy $E_{F}$ for normal and spin dissipation sources with different dissipation strength $\Gamma$, respectively.}
  \label {fig:2}
\end{figure}

\section{Numerical results and discussions}

\subsection{Thermal dissipation caused by normal or spin dissipation sources}

To better show the difference of normal and spin dissipation sources, we first study the influence of dissipation sources
on the quantized conductance plateaus in this section. Figure~\ref{fig:2}(a) gives the spin-resolved band structure of the HgTe/CdTe quantum well. From figure~\ref{fig:2}(a), we can see that the energy bands of spin-up and spin-down are degenerate and there are helical edge states in the bulk gap, which is a typical characteristic of the QSHE. Figures~\ref{fig:2}(b) and (c) show the conductance $G$ for normal and spin dissipation sources, respectively, as a function of Fermi energy $E_{F}$. One can see that in the absence of dissipation sources ($\Gamma=0$),
the conductance $G$ shows perfect quantized plateaus
with the plateau values of $ne^2/h$ ($n=2,4,6...$ is the number of intersections of the Fermi energy and the energy band in figure~\ref{fig:2}(a)). In the bulk gap,
the quantized conductance plateau of $2e^2/h$ appears due to the spin
degenerate helical edge states.
The other higher quantized plateaus in the bulk originate
from the transverse subbands of the HgTe/CdTe quantum well with finite width.
These phenomena are consistent with previous theoretical works \cite{JPCM.26.085301,PRL.103.036803}.
In the presence of dissipation sources, the conductance $G$ shows
similar behavior in the bulk for both normal and spin dissipation sources. That is, the conductance plateaus are destroyed and $G$ decreases
obviously with the increase of the dissipation strength $\Gamma$,
because dissipation sources enhance the scattering of electrons in the bulk, which is not protected by TRS.
In the following, we mainly focus on the quantum spin Hall regime
in the gap, where $G$ behaves quite differently depending
on the type of dissipation sources.
Due to TRS, the helical edge states with opposite spin
counterpropagate along the boundary of the device (figure~\ref{fig:1}(b)), producing a quantized conductance plateau of $2e^2/h$.
For normal dissipation sources, the quantized plateau remains well
with increasing $\Gamma$, see figure~\ref{fig:2}(b),
because carriers maintain their spin memories and backscattering occurs
unless the carriers are scattered from one boundary to another.
However, this is impossible since the two boundaries are far from each other in space.
Therefore, the quantized plateau of $2e^2/h$ is robust against a
large $\Gamma$ for normal dissipation sources.
On the other hand, for spin dissipation sources,
the quantized conductance plateau of $2e^2/h$ is destroyed and
decreases with increasing $\Gamma$, see figure~\ref{fig:2}(c). This is because the spin of carriers can be flipped by spin dissipation sources, thus backscattering occurs between the helical edge states on each boundary of the device (see figure~\ref{fig:1}(c)), leading to the failure of the quantized conductance plateau of $2e^2/h$. Since the dissipation strength $\Gamma$ is independent of energy $E$, the probability of backscattering between this pair of helical edge states is almost a constant in the gap. Therefore, the decreased conductance is still a plateau.

Next, we focus on the thermal dissipation of the helical edge states.
It is worth mentioning that in this paper we assume the perfect coupling
between the left (right) lead and the central region
thus no backscattering occurs at their interface. As a result, the injected electrons with spin-down (spin-up) from left (right) lead to the helical edge states of the central region keep equilibrium with the same chemical potential as that of the left (right) lead.
If there exists a barrier and scattering occurs at the interface,
the injected electrons with spin-up and spin-down propagating to the helical edge states are respectively in the non-equilibrium at the beginning, then the dissipation can occur regardless of the type of
the dissipation sources, which is similar to the dissipation occurring
at the chiral edge states in the IQHE system~\cite{PRB.104.115411}.

\begin{figure}
  \includegraphics[width=0.8\textwidth]{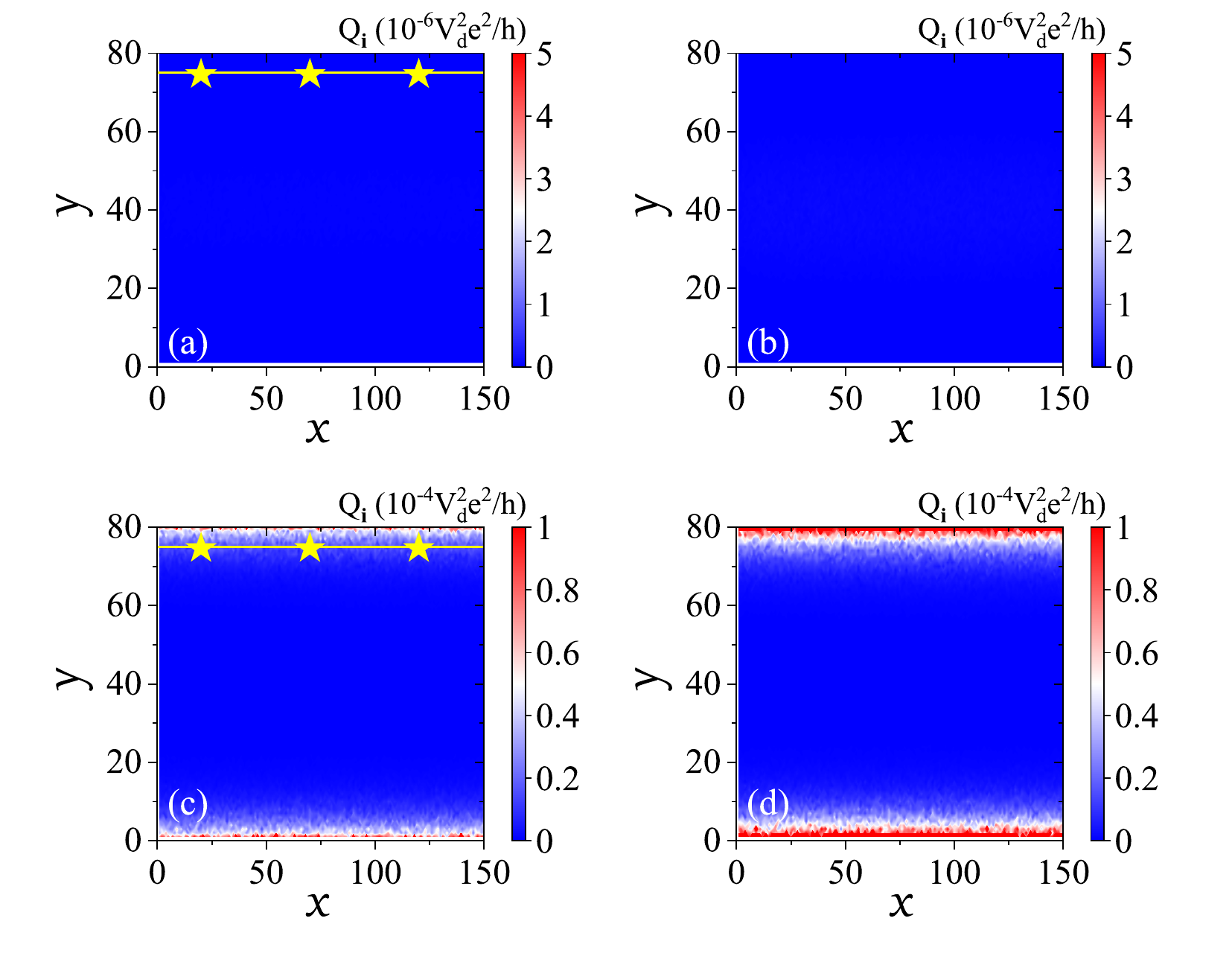}
  \centering
  \caption{Local heat generation $Q_{\bf i}$ vs lattice position $(x, y)$ with $E_{F}=0$. Considering normal dissipation sources in (a), (b), and spin dissipation sources in (c), (d). The dissipation strength $\Gamma=1$ meV in (a), (c) and 2 meV in (b), (d). Positions marked in (a), (c) are $y=75$ and $x=20, 70, 120$, respectively.}
  \label {fig:3}
\end{figure}

Figures~\ref{fig:3}(a) and (b) plot the local heat generation
$Q_{\bf i}$ versus the central lattice position $(x,y)$
for normal dissipation sources.
From figures~\ref{fig:3}(a) and (b), we can see
that no dissipation occurs at the helical edge states
regardless of the dissipation strength $\Gamma$.
For normal dissipation sources, TRS is preserved thus
the two edge states with opposite spin counterpropagate
along the boundary without backscattering (see figure~\ref{fig:1}(b)).
In this case, electrons with spin-up and spin-down always keep their own equilibrium during the propagation process,
so the helical edge states are dissipationless.
However, the results are completely different for spin dissipation sources. From figures~\ref{fig:3}(c) and (d), we can see that thermal dissipation occurs along the helical edge states
at the upper and lower boundaries of the central region for spin dissipation sources. In the presence of spin dissipation sources,
TRS is broken and spin-flip backscattering can occur between
the counterpropagating edge states (see figure~\ref{fig:1}(c)),
therefore a weak dissipation strength $\Gamma$ can lead to
the occurrence of thermal dissipation (see figure~\ref{fig:3}(c)). With the increase of $\Gamma$,
thermal dissipation will increase obviously due to the enhanced spin-flip backscattering (see figure~\ref{fig:3}(d)). Considering this point, if we can reduce the spin dissipation strength $\Gamma$ in the device, the thermal dissipation can be reduced obviously. Since the spin dissipation sources originate from magnetic impurities and nuclear spin, etc., one can choose a higher purity sample or the atoms with zero nuclear spin to reduce the influence of them as well as the thermal dissipation.

\subsection{Energy distribution of electrons}
To better understand the underlying mechanism of
thermal dissipation occurred in QSHE, we now analyze the energy distribution of
electrons at different positions. By using the Green's function,
the energy distribution of electrons with spin $\sigma$
at site ${\bf i}$ can be expressed as~\cite{PRB.104.115411}
\begin{equation}\label{eq:7}
F_{\sigma}({\bf i},E)=\frac{n_\sigma({\bf i},E)}
{\textrm{LDOS}_{\sigma}({\bf i},E)},
\end{equation}
where $n_\sigma({\bf i},E)=-\frac{i}{2\pi}\textrm{Tr}[{\bf G}_{\bf i\sigma,i\sigma}^<(E)]$
is the electron density per unit energy at the lattice site ${\bf i}$
and $\textrm{LDOS}_{\sigma}({\bf i},E)=-\frac{1}{\pi}\textrm{Tr}[\textmd{Im}{\bf G}_{\bf i\sigma,i\sigma}^{r}(E)]$
is the local density of states at the lattice site ${\bf i}$~\cite{Datta}.
According to the Keldysh equation, the lesser Green's function ${\bf G}_{{\bf i}\sigma,{\bf i}\sigma}^{<}$ and the lesser self-energy ${\bf \Sigma}_{p\sigma}^{<}$ are
\begin{eqnarray}\label{eq:8}
{\bf G}_{{\bf i}\sigma,{\bf i}\sigma}^{<}(E)&=&\sum_{p,\sigma'}
{\bf G}_{{\bf i\sigma},{\bf j}_p\sigma'}^{r}
{\bf \Sigma}_{p\sigma'}^{<} {\bf G}_{{\bf j}_p\sigma',{\bf i}\sigma}^{a}, \nonumber \\
{\bf \Sigma}_{p\sigma}^{<}(E)&=&-f_{p\sigma}(E)[{\bf \Sigma}_{p\sigma}^{r}(E)
-{\bf \Sigma}_{p\sigma}^{a}(E)].
\end{eqnarray}
Here ${\bf j}_p$ is the lattice site coupling to the lead $p$.

\begin{figure}
	\includegraphics[width=0.8\textwidth]{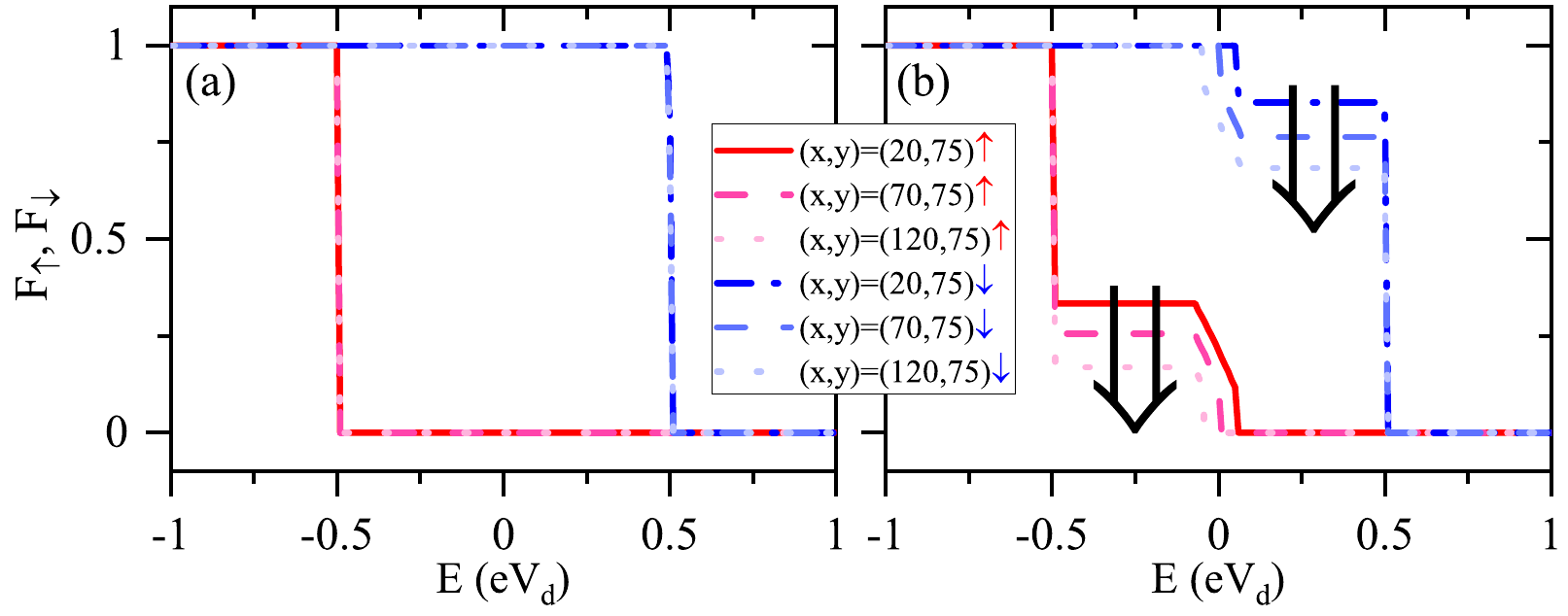}
	\centering
	\caption{ Distribution function $F_{\uparrow}$ (red lines) and $F_{\downarrow}$ (blue lines) versus energy $E$ for different positions given in figures~\ref{fig:3}(a) and (c), with (a) normal and (b) spin dissipation sources, respectively.}
	\label {fig:4}
\end{figure}

\begin{figure*}
	\includegraphics[width=0.8\textwidth]{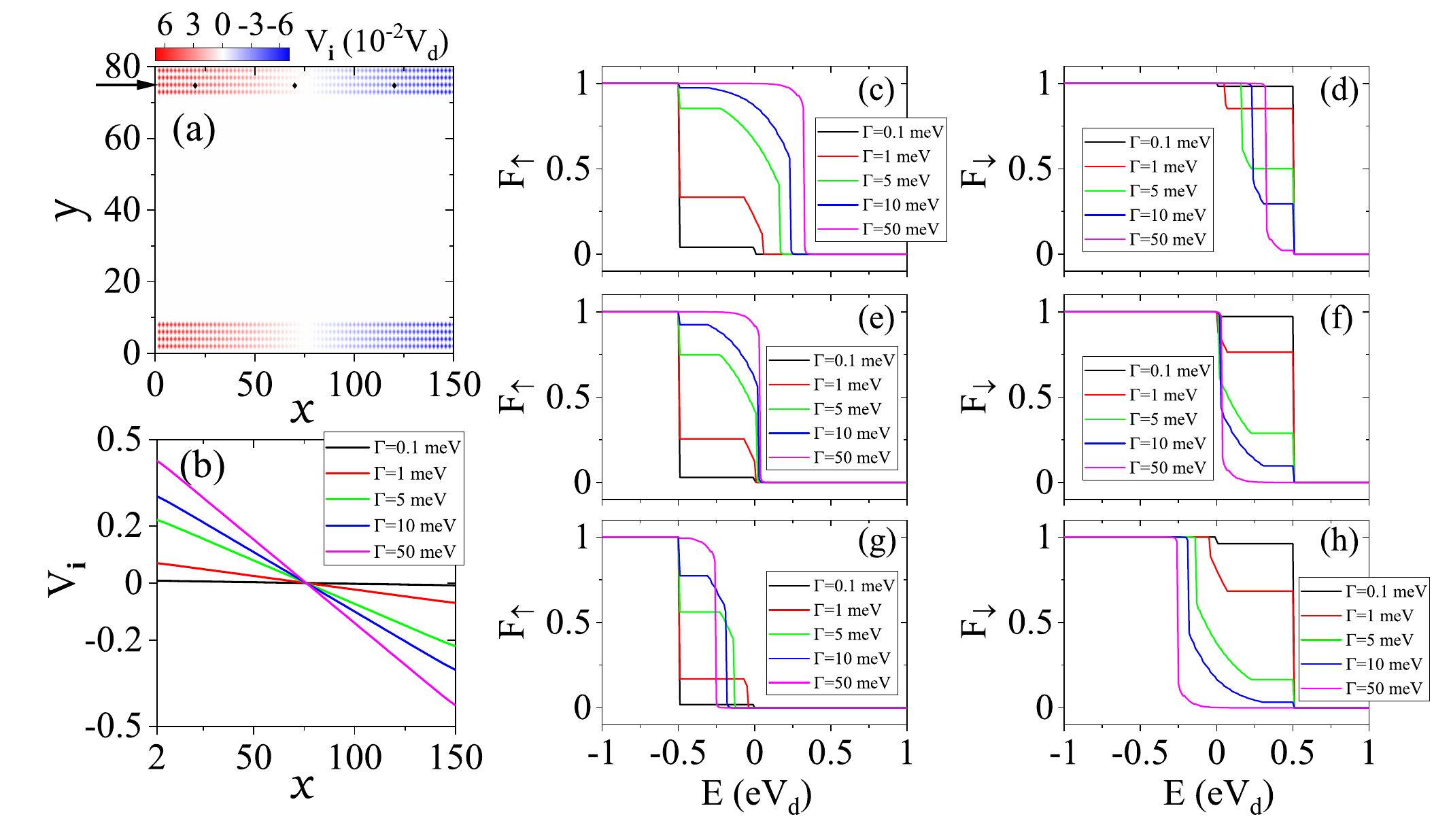}
	\centering
	\caption{ (a) The voltage $V_{\bf i}$ in virtual lead ${\bf i}$
	versus lattice position $(x,y)$ for spin dissipation sources with $N_v=8$ and $\Gamma=1$ meV. The black arrow marks the position with $y=75$.
	(b) The voltage $V_{\bf i}$ vs longitudinal position $x$ ($x=2, 4, ...N_x$) with transverse position $y=75$ for different spin dissipation strength $\Gamma$. (c)-(h) Distribution function $F_{\uparrow}$ and $F_{\downarrow}$ versus energy $E$ for different spin dissipation strength $\Gamma$ with transverse position $y=75$. The longitudinal positions $x=20$ in (c),(d), 70 in (e),(f), 120 in (g),(h), which are marked by small black diamonds in (a), respectively.}
	\label {fig:5}
\end{figure*}

When calculating the energy distribution,
virtual leads are added at the upper and lower boundaries of the central region.
Specifically, those lattices
with the $y$-coordinates $y=1,3,...N_v-1$ at the lower boundary
and $y=N_y-N_v+2, N_y-N_v+4,...N_y$ at the upper boundary
and the $x$-coordinates $x=2,4,...N_x$
are coupled to the virtual leads (see figure~\ref{fig:5}(a)),
and there are totally $\frac{N_vN_x}{2}$ virtual leads.
This is feasible since we only focus on the helical
edge states~\cite{JPCM.26.085301}.
The result is the same as virtual leads are added at the central region randomly with a certain probability $\eta$.
Figures~\ref{fig:4}(a) and (b) show the distribution $F_\sigma$ ($\sigma=\uparrow, \downarrow$) versus energy $E$ for different positions marked in figures~\ref{fig:3}(a) and (c), respectively.
From figure~\ref{fig:4}(a), we can find that $F_{\uparrow}=\theta(eV_R-E)$
and $F_{\downarrow}=\theta(eV_L-E)$ for all different positions,
which means that electrons on the helical edge states with spin-up and spin-down are always in their own equilibrium for normal dissipation sources.
Because the left and right leads are well coupled to the central region,
at the upper boundary, electrons with spin-down traveling clockwise
are all coming from the left lead with the same electrochemical
potential $eV_L$, and electrons with spin-up traveling anticlockwise
are all coming from the right lead with the same electrochemical
potential $eV_R$, see figure~\ref{fig:1}(b)~\cite{Datta}. In view of that, electrons with spin-up and spin-down are in their own equilibrium
and no spin-flip scattering occurs. Therefore, thermal dissipation cannot occur for normal
dissipation sources, as shown in figures~\ref{fig:3}(a) and (b).

However, the results are completely different for spin dissipation sources. From figure~\ref{fig:4}(b), we can see that $F_{\uparrow}$
and $F_{\downarrow}$ are non-equilibrium for different positions.
With the increase of longitudinal position $x$, the number of electrons
decreases in both spin-up and spin-down channels from left ($x=20$) to right ($x=120$), see the black arrows in figure~\ref{fig:4}(b).
At the upper boundary, electrons with spin-down come
from the left lead, and electrons with spin-up come from the right lead.
They are both in equilibrium at the beginning.
However, spin dissipation sources couple the spin-up and spin-down
subsystems together to induce spin-flip backscattering,
thus electrons will relax between the spin-up and spin-down channels.
That is to say, the counterpropagating edge states are constituted
partially by electrons originating from left lead and partially by electrons originating from right lead (see figure~\ref{fig:1}(c)).
Consequently, the energy distribution of electrons becomes non-equilibrium. Accompanying with the relax process,
electrons transfer the ordered electrical energy to environment (virtual leads) in the form of disordered thermal energy,
and the number of electrons with higher-energy decreases
but that with lower-energy increases.
Therefore, thermal dissipation occurs at the helical edge states,
as shown in figures~\ref{fig:3}(c) and (d).
More specifically, at the upper boundary of the sample,
electrons with spin-down (spin-up) have a higher (lower)
chemical potential $eV_L$ ($eV_R$).
When electrons with spin-down propagate clockwise along
the helical edge states, they will be flipped to spin-up continually,
which propagate anticlockwise along the helical edge states.
As a result, for spin-down channel, the number of electrons will decrease
along the spin propagation direction from left to right,
and conversely, for spin-up channel,
the number of electrons will increase
along the spin propagation direction from right to left (see figure~\ref{fig:4}(b)).

We then study the evolution of the energy distribution
with increasing dissipation strength $\Gamma$ for spin dissipation sources.
Figure~\ref{fig:5}(a) shows the voltage $V_{\bf i}$
in the virtual lead ${\bf i}$ versus lattice
position $(x,y)$ for spin dissipation sources.
Note that $V_{\bf i\downarrow} = V_{\bf i\uparrow} \equiv V_{\bf i} $
for spin dissipation sources. At the upper and lower boundaries of the central region,
we can see that the voltage decreases along the current
flow direction (see figure~\ref{fig:5}(a)).
In the presence of spin dissipation sources,
the spin-up and spin-down channels will couple to each other.
The spin-flip backscattering will occur and this
will lead to increasing resistance along the current flow direction.
Therefore, the voltage drops along the helical edge states
with increasing $x$~\cite{PRL.103.036803}.
Figure~\ref{fig:5}(b) shows the voltage $V_{\bf i}$
in the virtual lead ${\bf i}$ versus longitudinal position $x$
with $y=75$ (see the black arrow in figure~\ref{fig:5}(a)).
From figure~\ref{fig:5}(b), we can see that the voltage
decreases linearly as $x$ increases since
dissipation sources are uniformly added at the boundary.
When the dissipation strength $\Gamma$ is very small,
the voltage $V_{\bf i}$ is nearly zero as $x$ increases
(see $\Gamma=0.1$ meV in figure~\ref{fig:5}(b)).
For a very small $\Gamma$, the coupling between spin-down and
spin-up channels is very weak at the helical edge states.
In this case, electrons with spin-down (spin-up) have
the chemical potential about $eV_{L}$ ($eV_{R}$) propagating
clockwise (anticlockwise) along the upper boundary,
so the spin-up and spin-down subsystems are approximately
in their own equilibrium, see $\Gamma=0.1$ meV in figures~\ref{fig:5}(c)-(h).
Therefore, the voltage $V_{\bf i}$
can be viewed as the averaged voltage of $\frac{V_{L}+V_{R}}{2}$,
which is equal to zero.
That is to say,
the voltage drops equally at the left (right) lead-center
region interfaces due to the contact resistances,
but becomes flat across the helical edge states.
With the increase of the dissipation strength $\Gamma$,
the absolute value of slope for the curve $V_{\bf i}-x$ increases
(see figure~\ref{fig:5}(b)), which means the voltage $V_{\bf i}$ drops
faster with increasing $x$ for larger $\Gamma$.
This is because the spin-flip backscattering
is strengthened with increasing $\Gamma$, which leads to the increasing resistance at the helical edge states. When $\Gamma$ is large enough, spin-up and spin-down channels are completely coupled, and this evolves to the classical case with $V_{\bf i}|_{x=2}\approx V_{L}$ and $V_{\bf i}|_{x=N_x}\approx V_{R}$ (see $\Gamma=50$ meV in figure~\ref{fig:5}(b)).
Figures~\ref{fig:5}(c)-(h) show the energy distribution $F_{\sigma}$
versus energy $E$ for different dissipation strength $\Gamma$.
As we described above, electrons with spin-up and spin-down are
both in their own equilibrium with $F_{\uparrow}=\theta (eV_R-E)$
and $F_{\downarrow}=\theta (eV_L-E)$ for a very small $\Gamma$
(see the curve of $\Gamma=0.1$ meV in figures~\ref{fig:5}(c)-(h)).
As $\Gamma$ increases, spin-up and spin-down channels begin
to couple to each other, which leads to the occurrence of
thermal dissipation, and the distribution $F_{\sigma}$
at site ${\bf i}$ will become non-equilibrium,
as shown in figures~\ref{fig:5}(c)-(h) with $\Gamma=1$ meV.
By further increasing $\Gamma$, the thermal dissipation
will enhance continually and the non-equilibrium distribution $F_{\sigma}$ at site ${\bf i}$ evolves gradually to a new equilibrium distribution,
in which the subsystem of electrons with spin-up
keeps mutual equilibrium with that of the electrons with spin-down, and
the distribution $F_{\sigma}$ becomes independent of the spin index $\sigma$. When $\Gamma$ is large enough, the distribution $F_{\sigma}$ is
finally equilibrium with $F_{\sigma}|_{\bf i}=\theta(eV_{\bf i}-E)$.
For example, $V_{\bf i}|_{x=20,70,120}=0.326V_d$, $0.035V_d$, $-0.256V_d$
(see the curve with $\Gamma=50$ meV in figure~\ref{fig:5}(b)),
which corresponds to the distribution $F_{\sigma}|_{x=20,70,120}\approx \theta(0.33eV_{d}-E)$, $\theta(0.04eV_{d}-E)$, $\theta(-0.25eV_{d}-E)$, respectively (see the curve of $\Gamma=50$ meV in figures~\ref{fig:5}(c)-(h)).

\subsection{The influence of Rashba SOC and disorder}
\begin{figure}
	\includegraphics[width=0.8\textwidth]{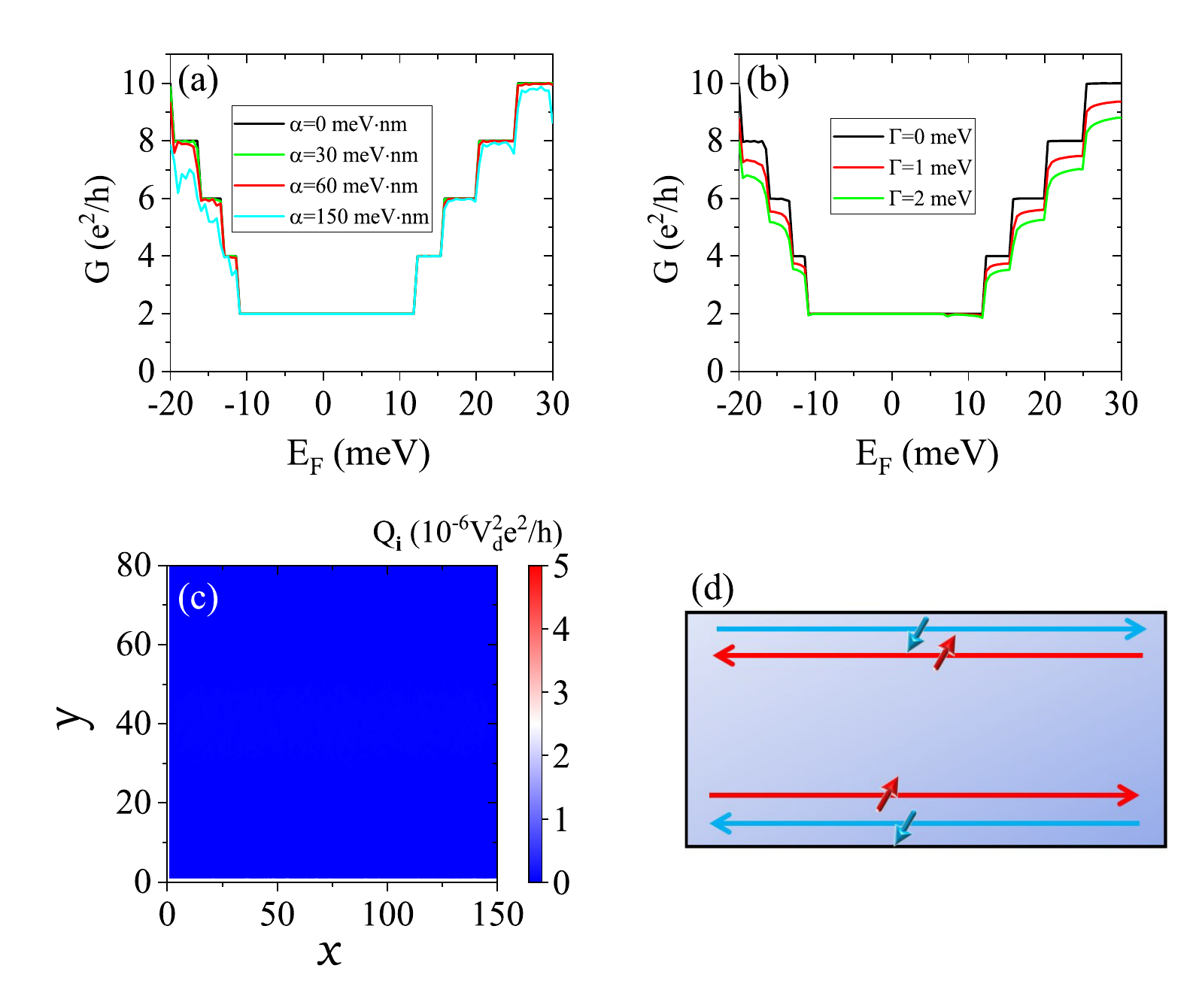}
	\centering
	\caption{ (a) Conductance $G$ versus Fermi energy $E_{F}$ for different Rashba SOC strength $\alpha$ with $\Gamma=0$. (b) Conductance $G$ versus Fermi energy $E_{F}$ in the presence of normal dissipation sources with $\alpha=30$ meV$\cdot$nm. (c) Local heat generation $Q_{\bf i}$ vs lattice position $(x, y)$ for normal dissipation sources with $\alpha=30$ meV$\cdot$nm and $\Gamma=1$ meV.
	(d) Propagation of helical edge states in the presence of Rashba SOC, spin-up and spin-down are twisted to an antiparallel direction by Rashba SOC.}
	\label {fig:6}
\end{figure}

The above results tell us that thermal dissipation can occur in QSHE for spin dissipation sources. Spin dissipation sources are a combination of lattice vibration and spin-flip process caused by magnetic impurities, nuclear spin, etc. Rashba SOC, which originates from unavoidable potential fluctuations, can also bring forth the spin-flip process~\cite{PRL.104.256804}. Therefore, if we combine Rashba SOC with normal dissipation sources, will the thermal dissipation also occur? To answer this question, we now study the thermal dissipation in QSHE with Rashba SOC in the central region. First, we give the conductance $G$ versus $E_{F}$ for different Rashba SOC strength $\alpha$, see figure~\ref{fig:6}(a). When $\alpha$ is small, the quantized conductance plateaus keep well (see $\alpha=30$ meV$\cdot$nm), since Rashba SOC still preserves TRS and backscattering between two TRS protected channels is forbidden. With the increase of $\alpha$, plateaus in the bulk are destroyed first and those in the hole area are destroyed more severely (see $\alpha=150$ meV$\cdot$nm). This arises from the fact that large $\alpha$ induces both the backscattering and strong electron-hole asymmetry in the bulk. Indeed, $\alpha=150$ meV$\cdot$nm is a very large value, because the Rashba SOC coefficient $\alpha$ in the inverted InAs/GaSb/AlSb quantum well is only $\alpha=71$ meV$\cdot$nm~\cite{PRL.100.236601}, while the Rashba SOC in HgTe/CdTe quantum well is much smaller than that in InAs/GaSb/AlSb quantum well. Therefore, the conductance plateau can survive well in the real clear sample of HgTe/CdTe quantum well. Figure~\ref{fig:6}(b) shows $G$ versus $E_{F}$ for normal dissipation sources in the presence of Rashba SOC. For the system with $\alpha=0$ (figure~\ref{fig:2}(b)) and $\alpha=30$ meV$\cdot$nm (figure~\ref{fig:6}(b)), the conductance $G$ is almost the same, indicating that the Rashba SOC has little effect on the two-terminal conductance, which is similar to previous studies of the four-terminal device~\cite{PRB.89.085309,PRL.95.136602}. Next, we focus on the thermal dissipation of the system in the presence of Rashba SOC. From figure~\ref{fig:6}(c), we can see that there is no thermal dissipation as well, although Rashba SOC induces the spin-flip process. This can be understood as follows. Without Rashba SOC, electrons with spin-up and spin-down are locked in two opposite directions along z-axis, see figure~\ref{fig:1}(b). After introducing the Rashba SOC, although electrons with spin-up and spin-down are twisted in a direction which deviates from z-axis, they must also be antiparallel and the edge states are still helical after twisting (figure~\ref{fig:6}(d)), 
because Rashba SOC cannot break TRS. At the same time, the normal dissipation sources are spin isotropic, which act on the electrons with spin-down and spin-up equally and also preserve the TRS, so no dissipation occurs in this case, as long as the Rashba SOC is not very strong to
close the bulk gap and destroy the helical edge states.

\begin{figure}
	\includegraphics[width=0.8\textwidth]{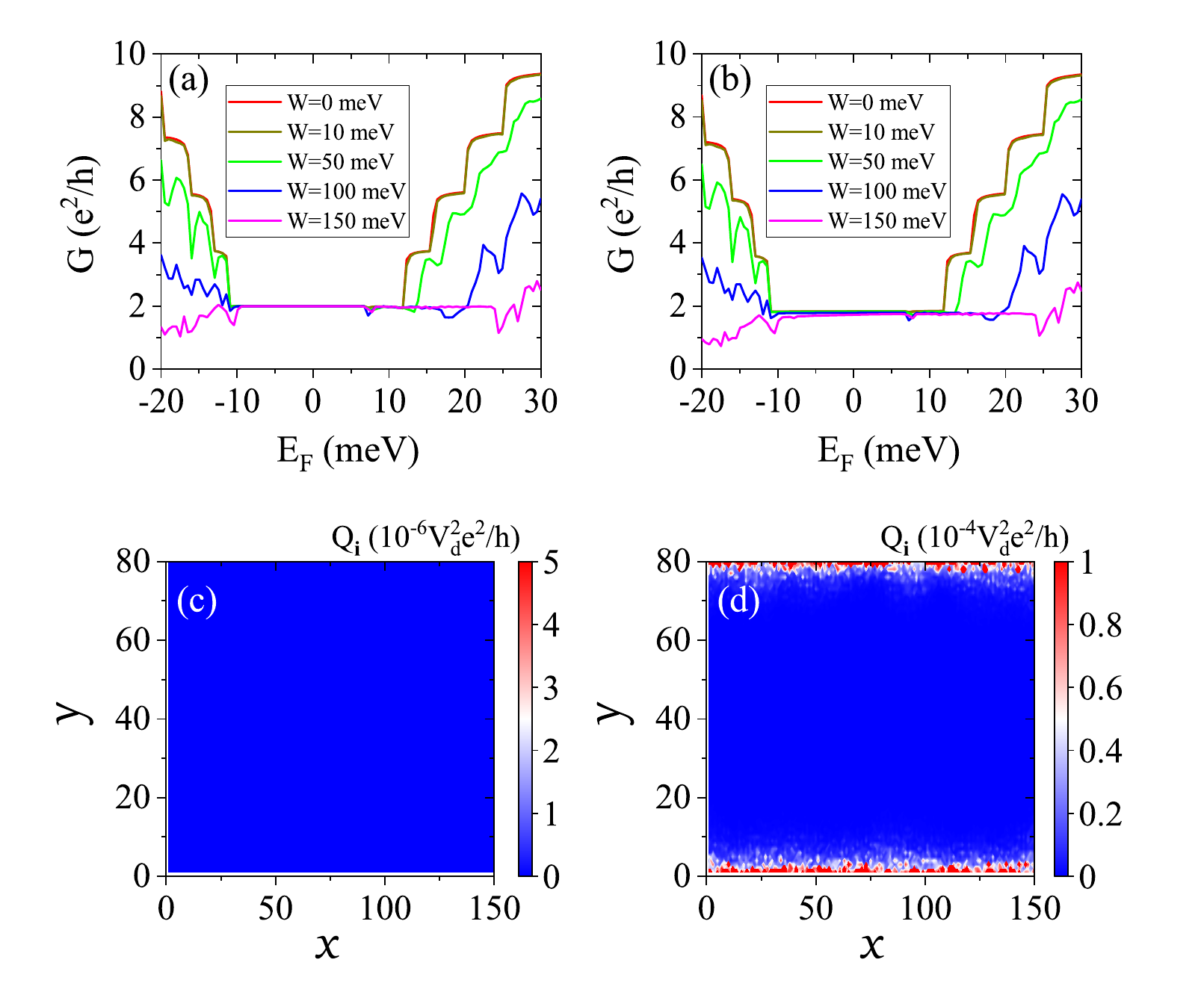}
	\centering
	\caption{(a), (b) Conductance $G$ versus Fermi energy $E_F$ for different disorder strength $W$ with (a) normal and (b) spin dissipation sources, respectively. The dissipation strength $\Gamma=1$ meV. (c), (d) Local heat generation $Q_{\bf i}$ vs lattice position $(x,y)$ in the presence of disorder with (c) normal and (d) spin dissipation sources, respectively. The dissipation strength $\Gamma=1$ meV and disorder strength $W=100$ meV.}
	\label {fig:7}
\end{figure}

In real samples, disorder is always inevitable. In the following, we study the effects of disorder on conductance $G$ and thermal dissipation $Q_{\bf i}$. Figures~\ref{fig:7}(a) and (b) show $G$ versus $E_{F}$ for normal and spin dissipation sources in the presence of disorder, respectively. For weak disorder ($W=10$ meV), $G$ is almost the same as $W=0$ (the red curve) for normal and spin dissipation sources, since weak disorder brings almost no scattering and extra spin-flip process. With the increase of $W$, the conductance plateaus in the bulk are destroyed first due to random scattering, but the plateau in the gap still keeps well, which means that the helical edge states are robust against moderate disorder. For relatively large $W$, the plateau of $2e^2/h$ can even be broaden (see $W=150$ meV). This indicates that disorder induces a new topological phase, and such a system is also known as the topological Anderson insulator~\cite{PRB.80.165316,PRL.102.136806,anref3}. Figures~\ref{fig:7}(c) and (d) show the thermal dissipation for both dissipation sources with large disorder strength $W=100$ meV. For normal dissipation sources, the helical edge states are dissipationless without disorder~(see figure~\ref{fig:3}(a)). From figure~\ref{fig:7}(c), we can see that there is still no dissipation in the presence of disorder, thus disorder has little effect on the occurrence of dissipation. But for spin dissipation sources, the helical edge states are dissipative (see figure~\ref{fig:3}(c)). Comparing with figures~\ref{fig:3}(c) and \ref{fig:7}(d), we can see that  the thermal dissipation is enhanced by disorder, since disorder enhances the probability of backscattering.

\section{Conclusion}
In summary, we have investigated the thermal dissipation processes
in QSHE based on HgTe/CdTe quantum well.
QSHE has a pair of helical edge states protected by TRS and such states are immune to normal backscattering theoretically,
thus providing great potential for future realization
of dissipationless conducting channels.
However, the inelastic scattering is still possible
in realistic samples for some special dissipation sources.
Here we study the influence of two kinds of dissipation sources on QSHE.
For normal dissipation sources, the helical edge states are dissipationless
with or without Rashba SOC, although Rashba SOC can induce spin-flip process. The core reason is that neither the normal dissipation sources
nor the Rashba SOC break TRS.
On the other hand, for spin dissipation sources,
TRS is broken and dissipation occurs along the helical edge states.
In particular, all electrons with spin-up and spin-down are in their own equilibrium, but the two subsystems are not equilibrium with each other.
For normal dissipation sources, the two subsystems are independent thus thermal dissipation will not occur.
However, spin dissipation sources can couple the two subsystems
together to induce voltage drop, and the distribution of electrons with spin-up and spin-down
will become non-equilibrium simultaneously, thus thermal dissipation will occur. With the increase of thermal dissipation, the energy distribution will evolve from non-equilibrium finally to mutual equilibrium for spin-up and spin-down subsystems.
The disorder effects are also considered. There is still no dissipation for normal dissipation sources after introducing disorder, but dissipation is enhanced for spin dissipation sources due to the enhanced backscattering. Our work provides clues to reduce thermal dissipation in the device based on the QSHIs.

\ack{This work was financially supported by NSF-China (Grants No.~11921005 and No.~12274466), the Innovation Program for Quantum Science and Technology (2021ZD0302403), the Strategic Priority Research Program of Chinese Academy of Sciences (Grant No. XDB28000000), and the Hunan Provincial Science Fund for Distinguished Young Scholars (Grant No.~2023JJ10058).}

\appendix
\section*{Appendix A: Detailed calculation process for the cases of the normal and spin dissipation sources}

\renewcommand{\theequation}{A\arabic{equation}}
\renewcommand{\thefigure}{A\arabic{figure}}

Let's show the detailed calculation process according to different boundary conditions brought by these two dissipation sources. Assuming that there are $n$ virtual leads in the system, from the Landauer-B\"{u}ttiker formula, i.e., equation~(\ref{eq:4}), we can obtain the following equations

\begin{tiny}
	\begin{equation}\label{eq:A1}
	\left(
		\begin{array}{c}
			I_{1\uparrow } \\
			I_{1\downarrow } \\
			I_{2\uparrow } \\
			I_{2\downarrow } \\
			\vdots \\
			I_{(n+2)\uparrow} \\
			I_{(n+2)\downarrow} \\
		\end{array}
	\right)
	=\frac{e^2}{h}
	\left(
		\begin{array}{ccccccccc}
			M_{1\uparrow} & 0 & -T_{1\uparrow,2\uparrow} & -T_{1\uparrow,2\downarrow} & \cdots & - T_{1\uparrow,(n+2)\uparrow} & T_{1\uparrow,(n+2)\downarrow} \\
			0 & M_{1\downarrow} & -T_{1\downarrow,2\uparrow} & -T_{1\downarrow,2\downarrow} & \cdots & -T_{1\downarrow,(n+2)\uparrow} & -T_{1\downarrow,(n+2)\downarrow} \\
			-T_{2\uparrow,1\uparrow} & -T_{2\uparrow,1\downarrow} & M_{2\uparrow} & 0 & \cdots & -T_{2\uparrow,(n+2)\uparrow} & -T_{2\uparrow,(n+2)\downarrow} \\
			-T_{2\downarrow,1\uparrow} & -T_{2\downarrow,1\downarrow} & 0 & M_{2\downarrow} & \cdots & -T_{2\downarrow,(n+2)\uparrow} & -T_{2\downarrow,(n+2)\downarrow} \\
			\vdots & \vdots & \vdots & \vdots & \ddots & \vdots & \vdots \\
			-T_{(n+2)\uparrow,1\uparrow} & -T_{(n+2)\uparrow,1\downarrow} & -T_{(n+2)\uparrow,2\uparrow} & -T_{(n+2)\uparrow,2\downarrow} & \cdots & M_{(n+2)\uparrow} & 0 \\
			-T_{(n+2)\downarrow,1\uparrow} & -T_{(n+2)\downarrow,1\downarrow} & -T_{(n+2)\downarrow,2\uparrow} & -T_{(n+2)\downarrow,2\downarrow} & \cdots & 0 & M_{(n+2)\downarrow}
		\end{array}
	\right)
	\left(
		\begin{array}{c}
			V_{1\uparrow } \\
			V_{1\downarrow } \\
			V_{2\uparrow } \\
			V_{2\downarrow } \\
			\vdots \\
			V_{(n+2)\uparrow} \\
			V_{(n+2)\downarrow} \\
		\end{array}
	\right)
	\end{equation}
	\end{tiny}\\
where $M_{p\sigma}=\sum_{q\neq p,\sigma'}T_{p\sigma,q\sigma'}$ with $p,q=1,2,\cdots,n+2$, and indices 1,2 represent the left and right leads, indices $3,4,\cdots,n+2$ represent the virtual leads ${\bf i}$. Define \\
$I_m=\begin{tiny}\begin{pmatrix} I_{1\uparrow } \\
	I_{1\downarrow } \\
	I_{2\uparrow } \\
	I_{2\downarrow } \end{pmatrix}\end{tiny}$,
$I_n=\begin{tiny}\begin{pmatrix} I_{3\uparrow } \\
		I_{3\downarrow } \\
		\vdots \\
		I_{(n+2)\uparrow } \\
		I_{(n+2)\downarrow } \end{pmatrix}\end{tiny}$,
$V_m=\begin{tiny}\begin{pmatrix} V_{1\uparrow }\\
			V_{1\downarrow } \\
			V_{2\uparrow } \\
			V_{2\downarrow } \end{pmatrix}\end{tiny}$,
$V_n=\begin{tiny}\begin{pmatrix} V_{3\uparrow}\\
				V_{3\downarrow } \\
				\vdots \\
				V_{(n+2)\uparrow } \\
				V_{(n+2)\downarrow } \end{pmatrix}\end{tiny}$.\\
$S_{11}=\begin{tiny}\begin{pmatrix} M_{1\uparrow} & 0 & -T_{1\uparrow,2\uparrow} & -T_{1\uparrow,2\downarrow} \\
0 & M_{1\downarrow} & -T_{1\downarrow,2\uparrow} & -T_{1\downarrow,2\downarrow} \\
-T_{2\uparrow,1\uparrow} & -T_{2\uparrow,1\downarrow} & M_{2\uparrow} & 0 \\
-T_{2\downarrow,1\uparrow} & -T_{2\downarrow,1\downarrow} & 0 & M_{2\downarrow}
\end{pmatrix}\end{tiny}$, $S_{12}=\begin{tiny}\begin{pmatrix} -T_{1\uparrow,3\uparrow} & -T_{1\uparrow,3\downarrow} & \cdots & -T_{1\uparrow,(n+2)\uparrow} & -T_{1\uparrow,(n+2)\downarrow}\\
-T_{1\downarrow,3\uparrow} & -T_{1\downarrow,3\downarrow} & \cdots & -T_{1\downarrow,(n+2)\uparrow} & -T_{1\downarrow,(n+2)\downarrow}\\
-T_{2\uparrow,3\uparrow} & -T_{2\uparrow,3\downarrow} & \cdots & -T_{2\uparrow,(n+2)\uparrow} & -T_{2\uparrow,(n+2)\downarrow}\\ -T_{2\downarrow,3\uparrow} & -T_{2\downarrow,3\downarrow} & \cdots & -T_{2\downarrow,(n+2)\uparrow} & -T_{2\downarrow,(n+2)\downarrow}
\end{pmatrix}\end{tiny}$,\\ 	
$S_{21}=\begin{tiny}\begin{pmatrix} -T_{3\uparrow,1\uparrow} & -T_{3\uparrow,1\downarrow} & -T_{3\uparrow,2\uparrow} & -T_{3\uparrow,2\downarrow} \\
-T_{3\downarrow,1\uparrow} & -T_{3\downarrow,1\downarrow} & -T_{3\downarrow,2\uparrow} & -T_{3\downarrow,2\downarrow} \\
\vdots & \vdots & \vdots & \vdots \\
-T_{(n+2)\uparrow,1\uparrow} & -T_{(n+2)\uparrow,1\downarrow} & -T_{(n+2)\uparrow,2\uparrow} & -T_{(n+2)\uparrow,2\downarrow} \\
-T_{(n+2)\downarrow,1\uparrow} & -T_{(n+2)\downarrow,1\downarrow} & -T_{(n+2)\uparrow,2\uparrow} & -T_{(n+2)\uparrow,2\downarrow}
\end{pmatrix}\end{tiny}$,\\
$S_{22}=\begin{tiny}\begin{pmatrix} M_{3\uparrow} & 0 & \cdots & -T_{3\uparrow,(n+2)\uparrow} & -T_{3\uparrow,(n+2)\downarrow} \\
0 & M_{3\downarrow} & \cdots & -T_{3\downarrow,(n+2)\uparrow} & -T_{3\downarrow,(n+2)\downarrow} \\
\vdots & \vdots & \ddots & \vdots & \vdots \\
-T_{(n+2)\uparrow,3\uparrow} & -T_{(n+2)\uparrow,3\downarrow} & \cdots & M_{(n+2)\uparrow} & 0\\
-T_{(n+2)\downarrow,3\uparrow} & -T_{(n+2)\downarrow,3\downarrow}  & \cdots & 0 & M_{(n+2)\downarrow}\end{pmatrix}\end{tiny}$,\\
then equation~(\ref{eq:A1}) can be rewritten by blocks as
	\begin{equation}\label{eq:A2}
	\left(
		\begin{array}{c}
			{I_m}_{(4\times1)} \\
			{I_n}_{(2n\times1)}
		\end{array}
	\right)
	=\frac{e^2}{h}
	\left(
		\begin{array}{cc}
			{S_{11}}_{(4\times 4)} & {S_{12}}_{(4\times2n)} \\
			{S_{21}}_{(2n\times 4)} & {S_{22}}_{(2n\times 2n)}
		\end{array}
	\right)
	\left(
		\begin{array}{c}
			{V_m}_{(4\times1)} \\
			{V_n}_{(2n\times1)}
		\end{array}
	\right).
	\end{equation}
Here the numbers $4\times 1$, $4\times 2n$, ... in the subscripts are
the dimensions of the vectors or matrices.
In equation~(\ref{eq:A2}), the transmission coefficient matrices $S$ and the voltage $V_m =\frac{V_d}{2}(1,1,-1,-1)^T$ are already known.
Combined with the different boundary conditions brought by the normal and spin dissipation sources, we can solve the rest currents $I_m$ and voltages $V_n$. In the following, let's show the detailed calculation process for normal and spin dissipation sources, respectively.

{\sl Calculation process for normal dissipation sources.}
For normal dissipation sources, we have $I_{\bf i\uparrow }=I_{\bf i\downarrow}=0$ with ${\bf i}=3,4,\cdots,n+2$. Substitute these conditions into equation~(\ref{eq:A2}), we can get that
\begin{eqnarray}\label{eq:A3}
	S_{11}V_m+S_{12}V_n&=&\frac{h}{e^2}I_m, \nonumber \\
	S_{21}V_m+S_{22}V_n&=&0.
\end{eqnarray}
By solving equation~(\ref{eq:A3}), we can get the voltages $V_{n}$ in the virtual leads and the currents $I_m$ in leads 1,2, i.e.,
\begin{eqnarray}\label{eq:A4}
V_n&=&-S_{22}^{-1}S_{21}V_{m}, \nonumber \\
I_m&=&\frac{e^2}{h}(S_{11}-S_{12}S_{22}^{-1}S_{21})V_m.
\end{eqnarray}

{\sl Calculation process for spin dissipation sources.}
For spin dissipation sources, from equation~(\ref{eq:A2}), we can obtain that
\begin{eqnarray}
	S_{11}V_m+S_{12}V_n&=&\frac{h}{e^2}I_m, \label{eq:A5}\\
	S_{21}V_m+S_{22}V_n&=&\frac{h}{e^2}I_n.  \label{eq:A6}
\end{eqnarray}
In this case, we have $I_{\bf i\uparrow }+I_{\bf i\downarrow }=0$ and $V_{\bf i\uparrow}=V_{\bf i\downarrow }$ for ${\bf i}=3,4,\cdots,n+2$. Let $V_{ p\uparrow}=V_{p\downarrow }=V_{p}$ with $p=1,2,\cdots,n+2$, the equation~(\ref{eq:A6}) can be rewritten as
\begin{tiny}
	\begin{equation}\label{eq:A7}
	\left(
		\begin{array}{c}
			-T_{3,1}V_1-T_{3,2}V_2+M_3V_3-\cdots-T_{3,n+1}V_{n+1}-T_{3,n+2}V_{n+2} \\
			-T_{4,1}V_1-T_{4,2}V_2-T_{4,3}V_3+\cdots-T_{4,n+1}V_{n+1}-T_{4,n+2}V_{n+2} \\
			\vdots \\
			-T_{n+1,1}V_1-T_{n+1,2}V_2-T_{n+1,3}V_3-\cdots+M_{n+1}V_{n+1}-T_{n+1,n+2}V_{n+2}\\
			-T_{n+2,1}V_1-T_{n+2,2}V_2-T_{n+2,3}V_3-\cdots-T_{n+2,n+1}V_{n+1}+M_{n+2}V_{n+2}
		\end{array}
	\right)
	=
	\left(
		\begin{array}{c}
			0 \\
			0\\
			\vdots \\
			0\\
			0
		\end{array}
	\right),
	\end{equation}
	\end{tiny}\\
where $T_{q,p}=\sum_{\sigma,\sigma'}T_{p\sigma,q\sigma'}$ and $M_{p}=\sum_{\sigma}M_{p\sigma}$. 
From the equation~(\ref{eq:A7}), we can obtain
the bias $V_{\bf i}$ in the virtual leads with ${\bf i}=3,4,\cdots,n+2$,
\begin{tiny}
	\begin{equation}
	\left(
		\begin{array}{c}
			V_{3} \\
			V_{4} \\
			\vdots \\
			V_{n+1} \\
			V_{n+2}
		\end{array}
	\right)
	=
	\left(
		\begin{array}{cccccc}
			M_3 & -T_{3,4} & \cdots & -T_{3,n+1} & -T_{3,n+2} \\
			-T_{4,3} & M_4  & \cdots & -T_{4,n+1,} &-T_{4,n+2,} \\
			\vdots \\
			-T_{n+1,3} & -T_{n+1,4} & \cdots & M_{n+1} &-T_{n+1,n+2} \\
			-T_{n+2,3} & -T_{n+2,4} & \cdots & -T_{n+2,n+1} & M_{n+2}
		\end{array}
	\right)^{-1}
	\left(
		\begin{array}{cc}
			T_{3,1} & T_{3,2} \\
			T_{4,1} &T_{4,2}  \\
			\vdots \\
			T_{n+1,1} & T_{n+1,2}  \\
			T_{n+2,1} & T_{n+2,2}
		\end{array}
	\right)
	\left(
		\begin{array}{c}
			V_1 \\
			V_2 \\
		\end{array}
	\right),
	\end{equation}
	\end{tiny}\\
with $V_1=\frac{V_d}{2}$ and $V_2=-\frac{V_d}{2}$.
After obtaining the bias $V_{\bf i}=V_{\bf i\uparrow}=V_{\bf i\downarrow}$
in the virtual leads,
substitute them into equation~(\ref{eq:A5}), the current $I_m$ can be solved straightforwardly.

\end{document}